\documentclass[letterpaper]{mc2025}

\usepackage{physics}
\usepackage{siunitx}
\usepackage{mhchem}
\usepackage{tikz}
\usetikzlibrary{shapes.geometric}
\usetikzlibrary{arrows.meta}
\usetikzlibrary{calc}
\usetikzlibrary{patterns}
\usetikzlibrary{fit}

\usepackage{nomencl} 
\makenomenclature

\title{Stochastic Method for Delayed Neutron Precursors Transport in Liquid Fuel}

\addAuthor[mathis.caprais@cea.fr]{Mathis Caprais}{1}

\addAffiliation{1}{Université Paris-Saclay, CEA, Service d'\'{E}tudes des Réacteurs et de Mathématiques Appliquées, Gif-sur-Yvette, 91191, France}

\addAuthor{Daniele Tomatis}{2}

\addAffiliation{2}{Newcleo Srl, Via Giuseppe Galliano 27, Torino, 10129, Italy}

\Abstract{%
This paper presents a novel stochastic method for modeling the transport of Delayed Neutron Precursors (DNPs) in liquid nuclear fuel. The method incorporates advection and diffusion effects into the Monte Carlo solution of the neutron balance equation by leveraging the Green’s function of the advection-diffusion-reaction (ADR) equation. For a 1D system, we demonstrate that the Green’s function can be interpreted as the Probability Density Function (PDF) of the position increment of a Brownian motion with drift. Using this interpretation, the position of DNPs is sampled via a time-of-flight process combined with a drift and diffusion model. The method is validated on a modified 1D rod problem, where results from the Monte Carlo implementation are compared against those obtained using a deterministic approach. The comparison confirms that the method accurately captures the impact of fuel velocity and diffusion on neutron flux. As expected, the fuel velocity shifts the neutron flux. Reactivity decreases as a function of speed while diffusion can counteract this decrease under certain conditions. While the current study is limited to 1D systems, the approach could be extended to higher dimensions and more complex geometries by replacing the Green’s function with the Stochastic Differential Equation (SDE) associated with the ADR equation.
}
\keywords{Monte-Carlo, Delayed Neutron Precursors, Green function, Liquid nuclear fuel}

\shortTitle{M&C 2025 Full Summary Template}

\authorHead{M. CAPRAIS}

\newcommand{\keff}{k_{\text{eff}}}
\newcommand{\sgn}{\mathop{\mathrm{sgn}}}

\DeclareSIUnit\pcm{pcm}

\begin{document}

\section{Introduction}
In this paper, we introduce a method for simulating the transport of Delayed Neutron Precursors (DNPs) in liquid nuclear fuel through Monte-Carlo simulation. Properly modeling DNP transport is essential in liquid fuel reactors because fission products are transported and diffused by the fuel flow, causing delayed neutrons to be emitted in part of the system that does not contribute to the chain reaction, thus often decreasing the reactivity of a neutron multiplying system.

The motion of DNPs within Monte-Carlo simulations has already been a subject of studies. Previous work coupled the Monte-Carlo code (SERPENT) with a deterministic DNPs balance equation solver \cite{aufiero2017monte, groth2021verification}. Given a precursors' source computed with the Monte-Carlo code, the DNPs concentration was calculated and positions of delayed neutrons were sampled from it. Directly tracking the DNPs has also been studied \cite{aufiero2017monte, aufiero2014calculating}, but in these cases, each DNP was tracked individually on-the-fly when sampled, resulting in high computational cost. Additionally, methods that directly track individual precursors within Monte Carlo simulations \cite{aufiero2017monte, aufiero2014calculating} fail to account for diffusion, which might be significant in liquid fuel reactors \cite{Wooten2018}.

We present a new method for DNPs advection-diffusion in the context of Monte-Carlo calculations. Diffusion and advection are here taken into account by deriving the Green's function of the advection-diffusion-reaction (ADR) equation, which, in 1D, is shown to be equivalent to the Probability Density Function (PDF) of the position increment of a Brownian motion with drift. While finding the general Green's function is not possible, generalizing DNPs transport by using pathlines for drift and Brownian motion for diffusion is possible.

Therefore, instead of sampling the Green's function of the DNPs balance equation, time-of-flights of the drifting precursors are sampled based on an exponential probability density function for the \(j\)-th DNP group. Their position is then updated by adding a drift term from advection and a Brownian motion with a Mean Square Displacement (MSD) proportional to the diffusion coefficient multiplied by the time-of-flight. The method is validated on a modified rod problem, which features non-zero fuel velocity, diffusivity coefficient and two regions. Results are compared against a deterministic calculation. Despite its simplicity, the problem provides insights on precursors' physics and for testing the method's capability to reproduce the effects of DNPs drift in liquid fuel reactors.

The paper is organized as follows Sec. \ref{sec:integral_neutron_balance} presents the pathline method and the integral formulation of the DNPs balance equation. This integral formulation is then applied to the general integral neutron balance equation to obtain an equation solely on the neutron flux. The test case is studied in Sec. \ref{sec:rod_problem}, alongside the numerical methods (both stochastic and deterministic) to solve the neutron balance equation. The results of both calculation methods are presented in Sec. \ref{sec:results}, and the paper concludes with a discussion of the results and future work in Sec. \ref{sec:conclusion}.
\section{Integral neutron balance equation with DNPs transport}\label{sec:integral_neutron_balance}
\subsection{Delayed Neutron Precursors equation}
The DNPs equation is a balance equation between advection, diffusion and decay of the DNPs, which is in its scaled form:
\begin{equation}
    \vb{u}^\prime\cdot\grad{C_j^\prime} - \frac{1}{\mathcal{A}}\div{D^\prime\grad{C_j^\prime}} + \frac{1}{\mathcal{B}_j} C_j^\prime = \frac{S_j^\prime}{\mathcal{B}_j}, \quad +\,\mathrm{b.c.},
    \label{eq:dnps_eq}
\end{equation}
with \(C_j^\prime = C_j / C_0\) the scaled concentration of the \(j\)-th group of DNPs (\(C_0\), \SI{}{\per\cubic\meter}), \(\vb{u}^\prime\) the scaled velocity field of the fluid (\(u_0\), \SI{}{\meter\per\second}), \(D^\prime\) the scaled diffusivity coefficient of the DNPs (\(D_0\), \SI{}{\square\meter\per\second}) and \(S_j^\prime\) the scaled source term of the \(j\)-th group of DNPs (\(S_0 \propto \lambda_j C_0\), \SI{}{\per\cubic\meter\per\second}) with \(\lambda_j\) the decay constant of the \(j\)-th group. In Eq. \eqref{eq:dnps_eq}, b.c. denotes boundary conditions. All the previous quantities with index zero denote reference values. The three dimensionless numbers are defined as \(\mathcal{A} = L u_0 / D_0\), \(\mathcal{B}_j = u_0 / L\lambda_j\) and \(\mathcal{C}_j = D_0 / \lambda_j L^2\), respectively the ratio of advection over diffusion, the ratio of advection over decay (i.e ratio of DNPs mean free path over typical length of the system, or the number of times the DNPs can go through the system before decaying) and the ratio of diffusion over decay (squared diffusion length over squared characteristic length).
\subsection{Integral formulation of the DNPs concentration}
Let \(G_j\) be the Green function (with b.c.) of the \(j\)-th DNPs equation Eq. \eqref{eq:dnps_eq}, the DNPs delayed activity can be expressed as an integral over the neutron flux \(\psi\),
\begin{equation}
    \lambda_j C_j(\vb{r}) = \int \dd{\vb{r^\prime}}G_j (\vb{r}^\prime, \vb{r})\beta_j\int_{E}\dd{E^\prime}\nu\qty(E^\prime)\Sigma_f (E^\prime)\phi(\vb{r}^\prime,E^\prime),
    \label{eq:integral_formulation}
\end{equation}
where \(\beta_j\) is the delayed neutron fraction of the \(j\)-th group of DNPs, \(\nu\) is the average number of neutrons produced per fission, \(\Sigma_f (E^\prime)\) is the fission cross-section and \(\phi(\vb{r}^\prime,E^\prime,\vb*{\Omega}^\prime)\) is the scalar neutron flux at position \(\vb{r}^\prime\), energy \(E^\prime\) and direction \(\vb*{\Omega}^\prime\). In Eq. \eqref{eq:integral_formulation}, \(G_j\) identifies as a DNP transport operator, expressing that a part of the neutron production is drifted by the velocity field of the fluid and diffusion from the position \(\vb{r}^\prime\) to the position \(\vb{r}\).
\subsection{Integral neutron balance equation}
The integral neutron balance equation for the collision density \(\varphi (\vb{r}, \vb*{\Omega}, E) = \Sigma_t \qty(\vb{r}, E)\psi\qty(\vb{r}, \vb*{\Omega}, E)\) can be written as:
\begin{equation}
\varphi (\vb{P}) = \int \dd{\vb{P}^\prime} K\qty(\vb{P}^\prime \to \vb{P}) \varphi \qty(\vb{P}^\prime), \qq{with} \vb{P} = \qty(\vb{r}, \vb*{\Omega}, E),
\label{eq:integral_neutron_balance}
\end{equation}
with \(K\qty(\vb{P}^\prime \to \vb{P}) = T\qty(\vb{r}^\prime \to \vb{r}, \vb*{\Omega}, E)C\qty(\vb{r}^\prime, \vb*{\Omega}^\prime, E^\prime \to \vb*{\Omega}, E)\) being the transport kernel expressing the production of \(\varphi\) at \(\vb{r}^\prime\) with \(C\) transported to \(\vb{r}\) by \(T\). With spatial transport of DNPs, neutron production is decomposed into a prompt and delayed parts, with the delayed production part being the integral formulation of the DNPs concentration, Eq. \eqref{eq:integral_formulation}. Therefore, the neutron equation becomes the neutron balance equation becomes:
\begin{equation*}
    \varphi (\vb{P}) = \int \dd{\vb{P}^\prime} T \qty(C_p + \sum_j G_j C_d^j ) \varphi \qty(\vb{P}^\prime),
\end{equation*}
with \(C_p\) the prompt production part of the collision kernel and \(C_d^j\) the delayed production part of the collision kernel for the \(j\)-th group of DNPs. The prompt production part of the transport kernel is:
\begin{equation}
    C_p \varphi \qty(\vb{P})= \chi_p (E) \int \dd{\vb{X}^\prime} (1-\beta (E^\prime))\nu (E^\prime)\frac{\Sigma_f (\vb{r}, E^\prime)}{\Sigma_t(\vb{r}, E^\prime) }\varphi (\vb{r}, \vb{X}^\prime), \qq{with} \vb{X} = \qty(\vb*{\Omega}, E),
    \label{eq:prompt_collision_kernel}
\end{equation}
where \(\chi_p\) is the prompt fission spectrum. The delayed part of the transport kernel is:
\begin{equation}
G_j C_d^j \varphi \qty(\vb{P})= \chi_d^j (E) \int \dd{\vb{r^\prime}}G_j (\vb{r}^\prime, \vb{r}) \int \dd{\vb{X}^\prime} \beta_j (E^\prime) \nu (E^\prime) \frac{\Sigma_f (\vb{r}^{\prime}, E^\prime)}{\Sigma_t(\vb{r}^{\prime}, E^\prime) } \varphi (\vb{r}^{\prime}, \vb{X}^\prime),
\label{eq:delayed_collision_kernel}
\end{equation}
where \(\chi_d^j\) is the delayed fission spectrum of the \(j\)-th group of DNPs. The main difference between the two production terms, the prompt production Eq. \eqref{eq:prompt_collision_kernel} and the delayed production Eq. \eqref{eq:delayed_collision_kernel} is that the delayed neutron colliding at \(\vb{r}\) have been produced at another position \(\vb{r}^{\prime}\) by the collision density \(\varphi\) and transported to \(\vb{r}\) by the transport operator \(G_j\).
\section{The rod problem}\label{sec:rod_problem}
The rod problem is a basic 1D model that differs from the typical 1D slab configuration by considering only two directions of flight: forward and backward. In this modified version, the rod problem consists of two distinct regions. A core with a non-zero fission cross-section and a heat sink with a zero fission cross-section. The system is closed (infinite and periodic along the \(x\)-axis, with fluid motion and diffusivity), meaning that leaving neutron and precursors enter the core again. The same slab representation was used by the authors in other studies to investigate the physics of liquid fuel \cite{Caprais2022,caprais2024study, PAZSIT2012206}. The setup is illustrated in Fig. \ref{fig:rod_problem}.
\begin{figure}[hbtp]
    \centering
    \begin{minipage}[c]{0.45\textwidth}
        \centering
        \begin{tikzpicture}
            \draw[thick] (0, 0.1) -- (0, -0.1);
            \draw[thick] (6, 0.1) -- (6, -0.1);
            \draw[thick] (3, 0.1) -- (3, -0.1);

            \draw[-] (0, 0) -- (3, 0) node[midway, above] {\(\Sigma_f \neq 0\)};
            \draw[-] (3, 0) -- (6, 0) node[midway, above] {\(\Sigma_f = 0\)};
            
            \node at (0, -0.4) {\(0\)};
            \node at (6, -0.4) {\(\mathcal{L} = \ell + \mathsf{H}\)};
            \node at (3, -0.4) {\(\ell\)};
            
        \end{tikzpicture}
        \caption{Layout of the 1D problem with the core and the recirculation loop.}
        \label{fig:rod_problem}
    \end{minipage}%
    \hspace{0.05\textwidth} 
    \begin{minipage}[c]{0.45\textwidth}
        \centering
        \vspace*{0.5cm} 
        \captionof{table}{Input data for the 1D problem \cite{PAZSIT2012206}.}
        \vspace{0.5cm}
        \begin{tabular}{cc}
            \hline
            \hline
            Parameter       & Value \\
            \hline
            \(\ell\)             & \SI{3}{\meter} \\
            \(\mathsf{H}\)          & \SI{3}{\meter} \\
            \(\nu\)          & \SI{2.45}{} \\
            \(\beta\)          & \SI{650}{\pcm} \\
            \(\lambda\) & \SI{1e-1}{\per\second} \\
            \(\Sigma_f\) & \SI{0.7}{\per\meter} \\
            \(\Sigma_s\) & \SI{98}{\per\meter} \\
            \(\Sigma_a\) & \SI{1.0}{\per\meter} \\
            \hline
        \end{tabular}
        \label{tab:physical_parameters}
    \end{minipage}
\end{figure}
The simulation parameters of the rod problem are given in Table \ref{tab:physical_parameters}. The cross-sections are chosen so that the reactor is critical without fuel motion. The scattering cross-section is sufficiently large to reproduce the physics of a molten salt composed of light nuclei, such as \ce{LiF}. This implies neutron diffusion behavior. Similar values are found in the literature \cite{Caprais2022, PAZSIT2012206}.
\subsection{Monte-Carlo solution}\label{subsection:monte_carlo}
The critical problem is solved using a non-analog Monte-Carlo method \cite{hébert2009applied}.
\subsection{Delayed Neutron Precursors transport}\label{subsection:dnps_transport}
\subsubsection{Green's function of the ADR equation}
In the rod problem, the DNPs balance equation is the 1D version of Eq. \eqref{eq:dnps_eq}. With constant coefficients, an integral form of the DNPs activity is obtained using Green's functions on the infinite domain. The equation for the Green's function is:
\begin{equation}
    -\mathcal{C} \dv[2]{G}{\hat{x}} + \mathcal{B} \dv{G}{\hat{x}} + G = \delta \qty(\hat{x} - \hat{x}^\prime), \qq{and} \lambda C = \int_{-\infty}^{+\infty}\dd{\hat{x}^\prime}G\qty(\hat{x} - \hat{x}^\prime)\beta\nu\Sigma_f (\hat{x}^\prime)\phi(\hat{x}^\prime),
    \label{eq:green_function}
\end{equation}
with \(\delta\) the Dirac distribution, \(\hat{x} = x/L\) and \(S\) the DNPs source. The solution of Eq. \eqref{eq:green_function} is a combination of two exponential, where the coefficients are determined by the continuity of the Green's function and by Eq. \eqref{eq:green_function} integrated over an infinitesimal interval around \(x = x^\prime\) (current continuity). The Green's function is:
\begin{equation}
    G\qty(\hat{x} - \hat{x}^\prime) = \frac{1}{\mathcal{B}\sqrt{1 + 4\mathcal{C}/\mathcal{B}^2}}\begin{cases}
        \exp(r_+ (\hat{x}^\prime - \hat{x}))\qq{for} \hat{x}^\prime < \hat{x},\\
        \exp(r_- (\hat{x}^\prime - \hat{x}))\qq{for} \hat{x}^\prime > \hat{x},
    \end{cases}
    \qq{with} r_{\pm} = \frac{\mathcal{B}}{2\mathcal{C}} \qty(1 \pm \sqrt{1 + 4\frac{\mathcal{C}}{\mathcal{B}^2}}),
    \label{eq:green_function_solution}
\end{equation}
where \(r_\pm\) are the roots of the characteristic equation of Eq. \eqref{eq:green_function}. The Green's function can be interpreted as the probability density function of the DNP position increment. The Green's function defined by Eq. \eqref{eq:green_function_solution} is strictly positive and is normalized to unity.
\subsubsection{Sampling of the delayed neutron position}
Instead of directly sampling the distribution defined by Eq. \eqref{eq:green_function_solution}, the motion of a DNP is modeled as a Brownian motion with a drift and a diffusion coefficient. Both approaches are equivalent, but the position of the DNP is easier to sample. When the time-of-flight of the precursor is sampled according to an exponential law with parameter \(\lambda\), the position increment follows a normal distribution centered around the drift of the DNP,
\begin{equation}
    \Delta x \mid \tau \sim \mathcal{N}(u\tau, 2D\tau).
    \label{eq:dnps_position}
\end{equation}
The Mean Square Displacement (MSD) of the delayed neutron is \(\expval{\Delta x^2} = 2D\tau\), which is the MSD of a Brownian motion after a time \(\tau\) in 1D. Without advection, the position increment follows a normal distribution centered around zero, which is consistent with the fact that the diffusion operator of Eq. \eqref{eq:dnps_eq} is symmetric under the parity transformation \(x\to -x\) (left and right directions are equally probable). The PDF of the position increment is obtained by integrating out the time-of-flight \(\tau\),
\begin{equation}
    g(\hat{x}) = \int_{0}^{+\infty}\dd{\tau}f\qty(\hat{x}\mid\tau)p\qty(\tau), \quad f\qty(\hat{x}\mid\tau) = \frac{1}{\sqrt{4\pi D\tau}}\exp\qty(-\frac{\qty(\hat{x}L - u\tau)^2}{4D\tau}) \qq{and} p\qty(\tau) = \lambda\exp(-\lambda\tau).
    \label{eq:adv_diff_pdf}
\end{equation}
The argument of the exponential in Eq. \eqref{eq:adv_diff_pdf} can be developed and factorized to obtain an integral of the form \(\int_{0}^{+\infty}\dd{\tau}\tau^{-1/2}\exp(-\frac{a}{\tau}-b\tau)\). This integral is known and can be expressed in terms modified Bessel functions. Marginalizing over the time-of-flight, the PDF of the position increment is:
\begin{equation}
    g(\hat{x}) = \frac{1}{\mathcal{B}\sqrt{1+4\frac{\mathcal{C}}{\mathcal{B}^2}}}\exp(-\hat{x}\frac{\mathcal{B}}{2\mathcal{C}}\qty(\sgn{\hat{x}}\sqrt{1+4\frac{\mathcal{C}}{\mathcal{B}^2}} - 1)), \qq{and} \int_{-\infty}^{+\infty}\dd{\hat{x}}g(\hat{x}) = 1.
    \label{eq:adv_diff_pdf_final}
\end{equation}
The PDF defined by Eq. \eqref{eq:adv_diff_pdf_final} is actually the Green's function of the ADR (with \(\hat{x}\equiv \hat{x}^\prime - \hat{x}\)), Eq. \eqref{eq:dnps_eq}. Because both PDFs are the same, the position increment will be sampled in the next sections according to Eq. \eqref{eq:dnps_position} by first sampling a time-of-flight and then a shifted normal distribution. This is preferred over calculating numerically the cumulative distribution of Eq. \eqref{eq:green_function_solution}, sampling a random number with a uniform PDF, and calculating the position increment.

The mean value of the position increment is unaffected by diffusion, as \(\expval{\Delta x} = u / \lambda\). However, its variance is increased by the diffusion process, \(\expval{\Delta x^2} = 2\qty(\mathcal{B}^2 + \mathcal{C})L^2\). The entropy of the PDF is \(S = 1 + 1/2\log(\mathcal{B}^2 L^2 + 4\mathcal{C}L^2)\), meaning that if the velocity or the diffusion coefficient is increased, DNPs are more spread out in space.

Asymptotic limits of the PDF can be derived by taking the limits of the dimensionless numbers \(\mathcal{B}\) and \(\mathcal{C}\). The diffusion-free limit is when \(\mathcal{C}\to 0\), the PDF reduces to an exponential distribution, \(u / \lambda \exp(-\lambda x / u)\). Conversely, in the advection-free limit (\(\mathcal{B}\to 0\)), the PDF tends to the PDF of a Laplace distribution with zero mean. This PDF is symmetric (diffusion is invariant under parity transformation) and its scale parameter \(\sqrt{D/\lambda}\) represent the distance a precursor travels by diffusion before decaying. Long-lived precursor have a larger scale parameter \(\propto \sqrt{T_{1/2}}\), meaning that they are more affected by diffusion. Finally, the ``no motion'' limit is obtained with \(\mathcal{C}\to 0\) and \(\mathcal{B}\to 0\). The PDF tends to a Dirac distribution centered at the origin. This is the expected behavior as the precursors are not transported by the velocity field nor by diffusion, which is consistent with Eq. \eqref{eq:green_function} with \(\mathcal{C} = \mathcal{B} = 0\).
\subsection{Deterministic solution}
To validate the Monte-Carlo implementation, with the drift of DNPs, we perform a deterministic calculation. The neutron balance equation for the one-velocity rod problem is given by:
\begin{equation}
    \pm \pdv{\psi_\pm}{x} + \Sigma_t \psi_\pm = \frac{\Sigma_s}{2}\qty(\psi_{+} + \psi_{-}) + \qty(1 - \beta)\frac{\nu\Sigma_f}{2\keff}\qty(\psi_{+} + \psi_{-}) + \frac{\lambda}{2}C,
    \label{eq:neutron_balance}
\end{equation}
where \(\psi_{\pm}\) (\SI{}{\per\square\meter\per\second}) represents the angular neutron flux in the left and right directions, \(\Sigma_t\) (\SI{}{\per\meter}) is the total cross-section, \(\Sigma_s\) (\SI{}{\per\meter}) is the scattering cross-section, \(\Sigma_f\) (\SI{}{\per\meter}) is the fission cross-section, and \(\beta\) is the delayed neutron fraction. \(\keff\) is defined as the eigenvalue which avoids trivial solutions. By summing Eq. \eqref{eq:neutron_balance} for the left and right directions, we derive a diffusion equation for the scalar flux \(\phi = \psi_{+} + \psi_{-}\). Conversely, by subtracting Eq. \eqref{eq:neutron_balance} in both directions, we obtain the equation for the neutron current \(J = \psi_{+} - \psi_{-}\) that provides a Fick's law relation with the scalar flux. After substitution of the expression for the current, the resulting equation for the neutron scalar flux is:
\begin{equation}
    -\pdv{}{x}D\pdv{\phi}{x} + \Sigma_t \phi = \Sigma_s \phi + \qty(1 - \beta)\frac{\nu\Sigma_f}{\keff} \phi + \lambda C, \qq{with} D = 1 / \Sigma_t,\\
    \label{eq:neutron_scalar_flux}
\end{equation}
solved together with the DNP balance equation, Eq. \eqref{eq:dnps_eq}, with a source divided by \(\keff\). Eq. \eqref{eq:neutron_scalar_flux} is solved using the finite volume (FV) method with Fick's currents at the cell interfaces. Using a diffusion approximation is appropriate because \(\Sigma_s \simeq \Sigma_t\), Table \ref{tab:physical_parameters}. Periodic boundary conditions for the scalar flux, neutron current, and DNPs concentration are imposed at the problem's boundaries. The solution of the eigenvalue problem defined by Eq. \eqref{eq:neutron_scalar_flux} and Eq. \eqref{eq:dnps_eq} is found using the \texttt{scipy.sparse.linalg.eigs} method that wraps the ARPACK library (Scipy v. 1.14.1).
\section{Results}\label{sec:results}
\subsection{Convergence study \& comparison with deterministic calculation}\label{subsection:convergence}
To ensure the accuracy of the diffusion calculation, a mesh convergence study was conducted (with no advection nor diffusion). The cell count in the system was adjusted until the reactivity, determined in a steady-state configuration with tolerance of \(\SI{0.1}{\pcm}\). Convergence was achieved with \(\SI{1e4}{}\) mesh cells. The computed effective multiplication factor is \(\keff = \SI{1.003115}{}\). Similarly, a convergence study was carried out for the Monte-Carlo calculation. The parameters varied include the number of particles per batch, the number of active batches, and the number of inactive batches. Convergence was attained with \(\SI{2e5}{}\) particles per batch, \(\SI{4e3}{}\) active batches, and \(\SI{1.5e3}{}\) inactive batches. Given the highly diffusive nature of the fuel, a Russian Roulette threshold of \(\SI{0.8}{}\) and a survival weight of \(\SI{1.0}{}\) were selected. The effective multiplication factor was calculated to be \(\keff = \SI{1.00313 \pm 0.00005}{}\), which is in excellent agreement with the deterministic calculation, as it falls within the standard deviation of the Monte-Carlo result. The implementation of DNPs diffusion was validated with a large value of \(\mathcal{C} = \SI{1e5}{}\), corresponding to a large diffusion coefficient. The effective multiplication factor is \(\keff = \SI{0.99948 \pm 0.00005}{}\), which is in agreement with the deterministic calculation (\(\keff = \SI{0.999441}{}\)).
\subsection{Reactivity as a function of the fuel velocity}
\begin{figure}[hbtp]
    \centering
    \includegraphics[width=0.7\textwidth]{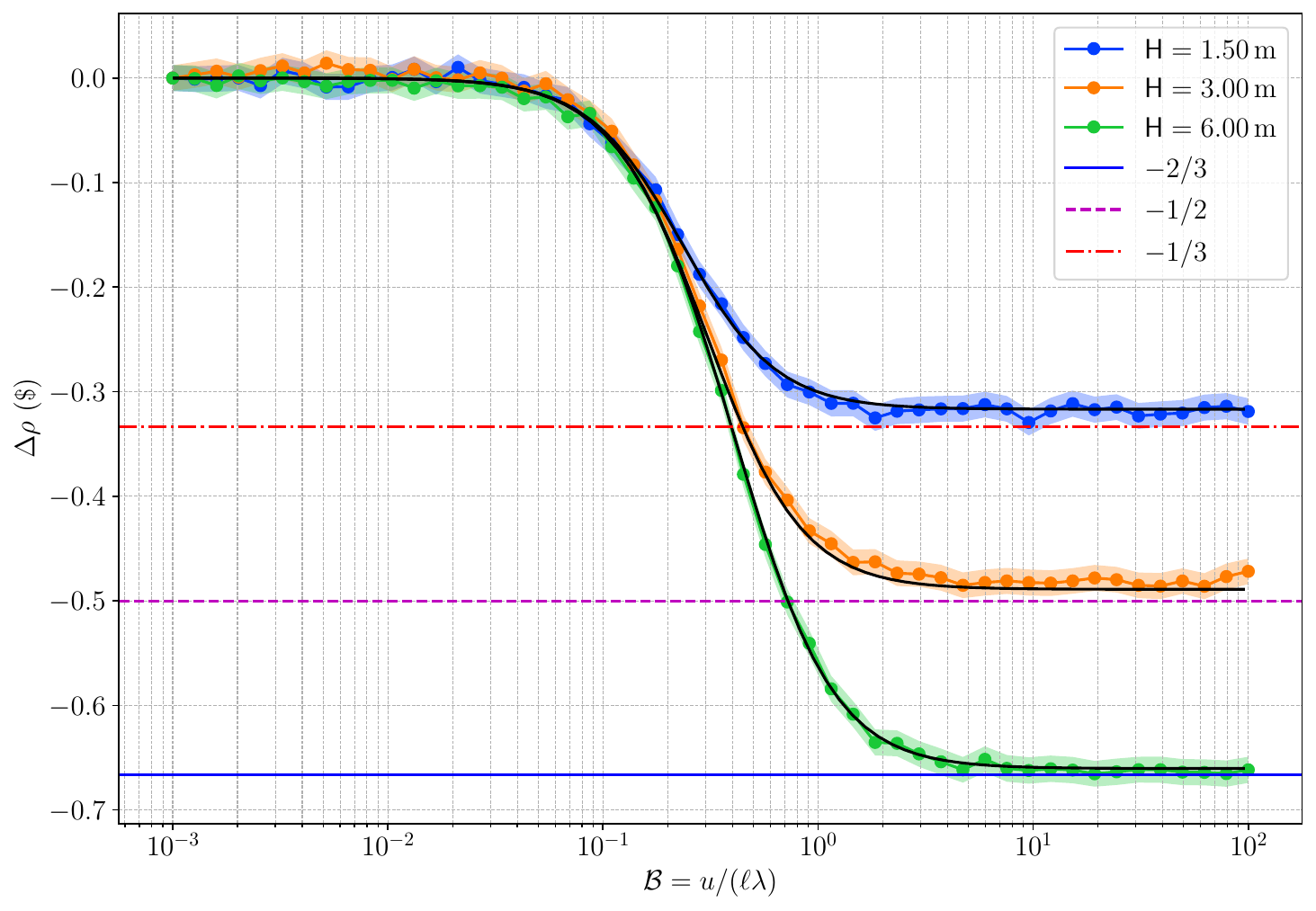}
    \caption{Reactivity as a function \(\mathcal{B}\) for different sizes of recirculation zones, with two standard deviations for the values of reactivity.}
    \label{fig:reactivity}
\end{figure}

The dimensionless number \(\mathcal{B}\) is varied to study the evolution of reactivity as a function of fuel velocity. Low values of \(\mathcal{B}\) correspond to a solid fuel reactor, while high values correspond to a reactor with high flow rates. The reactivity difference between the steady-state fuel configuration and the moving fuel configuration is calculated as \(\ln(\keff / k_{\mathrm{stat}})\) and is rescaled by a \(\beta\) factor to obtain the reactivity difference in percentage of the delayed fraction. The calculation is repeated for different sizes of the recirculation loop to highlight its impact on reactivity loss. This calculation is performed for both diffusion and Monte-Carlo methods, and the results are presented in Fig. \ref{fig:reactivity}.
The reactivity difference between the steady-state fuel configuration and the high flow rate configuration depends on the size of the outer loop. Longer recirculation loops result in a greater reactivity difference because more DNPs decay in this part of the core. The deterministic calculation (black solid lines in Fig. \ref{fig:reactivity}) lies within two standard deviations of the Monte-Carlo calculation, indicating good agreement between the two methods.
\subsection{Entropy}
\begin{figure}[h!]
    \centering
    \includegraphics[width=0.7\textwidth]{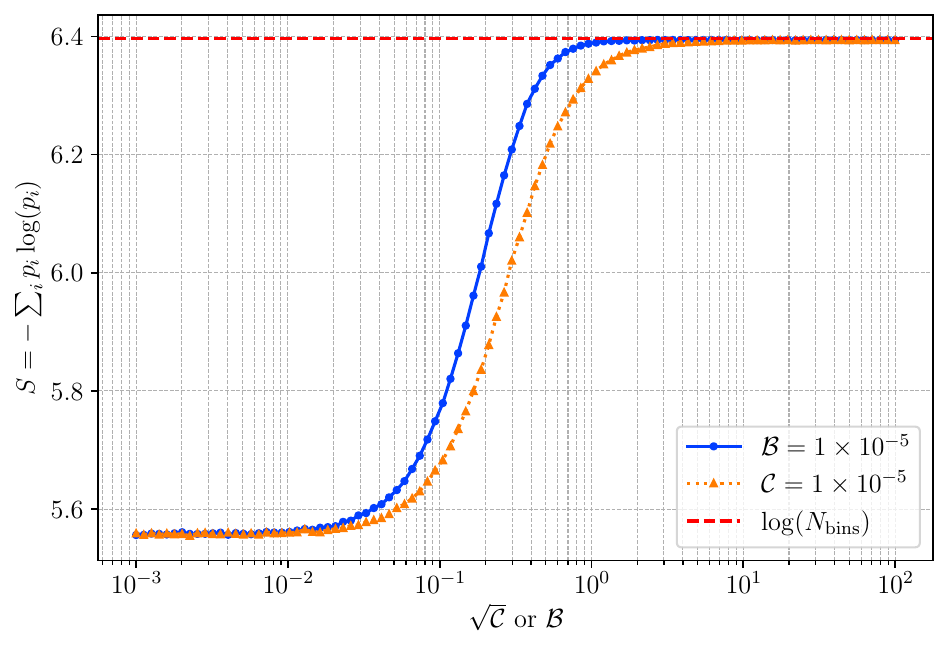}
    \caption{Entropy of the DNPs decaying positions as a function of \(\mathcal{B}\).}
    \label{fig:entropy}
\end{figure}

In Fig. \ref{fig:reactivity}, the change of reactivity lies between values of \(\mathcal{B} \in [10^{-1}, 10]\), which corresponds to DNPs mean free path of \SI{0.3}{\meter} to \SI{3}{\meter}. When the DNP mean free path is small compared to the size of the system, the effect of DNPs drift is negligible. In the opposite case, when the DNP mean free path is large compared to the size of the system, DNPs are diluted within the system. 
This homogenization process can be quantified by calculating the entropy of the delayed activity over spatial bins in the system. From a starting position randomly sampled from \(x_0 = \frac{\ell}{\pi} \arccos(1 - 2\eta)\), with \(\eta \sim \mathcal{U}(0,1)\) which represents a source of DNPs following a sinusoidal distribution within the core (\(S \propto \sin(\pi x/\ell)\)). For different values of the advection-reaction number \(\mathcal{B}\), \(N = \SI{2e5}{}\) DNPs are sampled, and their decay position is calculated using Eq. \eqref{eq:dnps_position}. The decaying positions are then scored in \SI{6e2}{} spatial bins. The homogenization process can be quantified by calculating the entropy of the DNPs decaying positions as a function of \(\mathcal{B}\) or \(\sqrt{\mathcal{C}}\), which is presented in Fig. \ref{fig:entropy}. Entropy is plotted as a function of \(\sqrt{\mathcal{C}}\) instead of \(\mathcal{C}\) because it represents the squared diffusion length over the squared characteristic length.
The maximum obtainable entropy, \(\log(N_{\text{bins}})\) is also displayed in Fig. \ref{fig:entropy}. The entropy plateau is reached with a value of \(\mathcal{B}\) which coincides with the values required to reach the reactivity difference plateau. High values of \(\mathcal{C}\) or \(\mathcal{B}\) maximize the entropy, meaning that every position in the system becomes equiprobable. Diffusion appears to homogenized DNPs before advection.
\subsection{Flux shift in the core}
\begin{figure}[h!]
    \centering
    \includegraphics[width=0.7\textwidth]{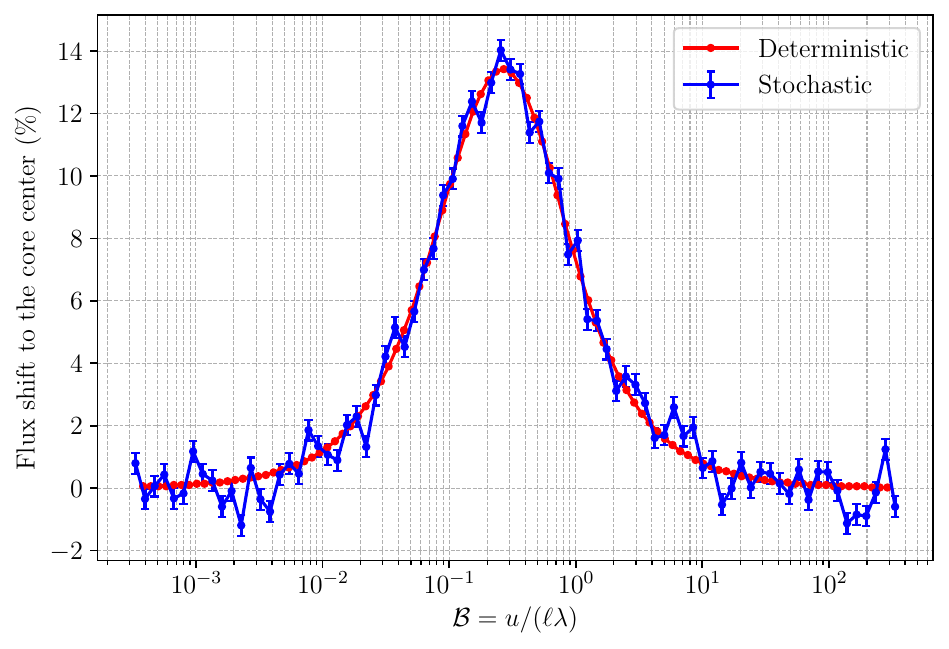}
    \caption{Flux shift in the core as a function of the fuel velocity.}
    \label{fig:flux_shift}
\end{figure}

The shift in the neutron flux can be calculated as a function of the advection-reaction number and compared to the deterministic calculation. The results are presented in Fig. \ref{fig:flux_shift}.
The neutron flux appears to be shifting for both stochastic and deterministic calculation for \(\mathcal{B} \in [10^{-1}, 10^0]\). This transition between fixed fuel and moving fuel is also present in the same range of advection-reaction number for the reactivity difference, Fig. \ref{fig:reactivity} and entropy of decay position of DNPs, Fig. \ref{fig:entropy}. The neutron flux is shifted to a maximum value of \SI{14}{\percent} of the half core length (roughly \SI{20}{\centi\meter}) and returns to its original position as the fuel velocity increases.
\subsection{Reactivity as a function of diffusion-reaction}
\begin{figure}[h!]
    \centering
    \includegraphics[width=0.7\textwidth]{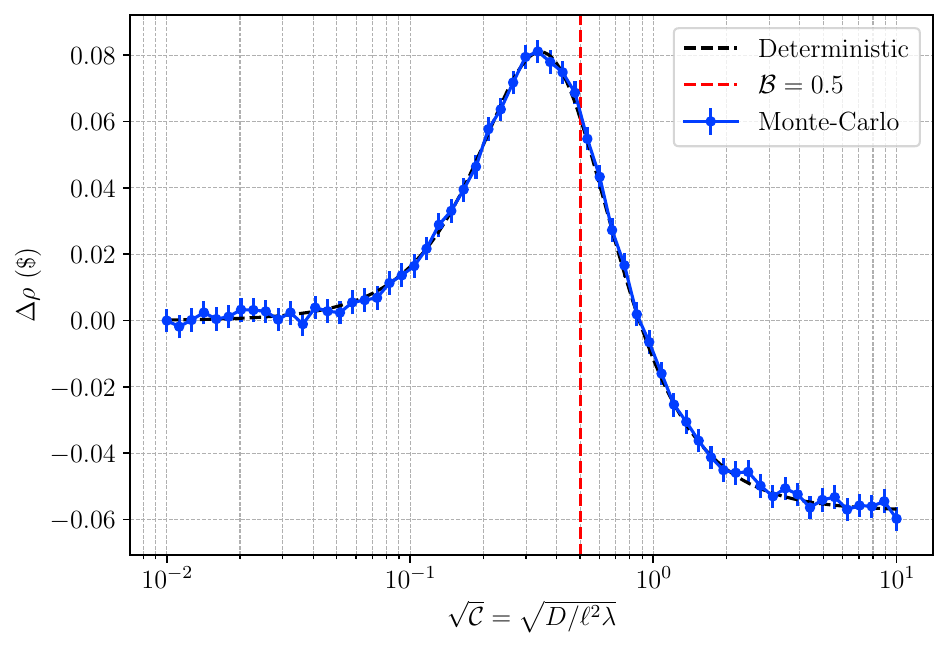}
    \caption{Reactivity difference between advection-only and advection-diffusion as a function of the diffusion-reaction number for a fixed advection-reaction number.}
    \label{fig:reactivity_diffusion}
\end{figure}

The diffusion-reaction number \(\mathcal{C}\) is varied while keeping the advection-reaction number constant (\(\mathcal{B}=\SI{0.5}{}\)). The reactivity difference between the advection-only configuration and the advection-diffusion configuration is calculated as \(\ln(\keff / k_{\mathrm{adv}})\). The results are presented in Fig. \ref{fig:reactivity_diffusion}. With the value of \(\beta\) chosen in Table \ref{tab:physical_parameters}, the reactivity surge is of the order of \SI{20}{\pcm}, which requires a lot of particles to be accurately calculated. Therefore, the value of \(\beta\) is increased to \SI{2e-2}{} to better observe the phenomenon.
Contrary to advection-only results, the reactivity difference between the advection-only and advection-diffusion configurations first increases with the diffusion-reaction number. This increase is due to diffusion moving back DNPs towards the core, increasing the reactivity. This phenomenon occurs roughly when the diffusion characteristic length reaches the DNPs advection mean free path, \(u / \lambda \propto \sqrt{D/\lambda}\) so \(\mathcal{C} \propto \mathcal{B}^2\) which is consistent with the value of \(\mathcal{B}\) taken for the calculation. The reactivity difference then decreases as the diffusion-reaction number increases, as the diffusion term dilutes the DNPs within the system.
\section{Conclusion}\label{sec:conclusion}
In this work, we showed a new method to account for the drift of Delayed Neutron Precursors (DNPs) within liquid nuclear fuel in the context of Monte-Carlo simulations. This drift introduces an additional integral in space in the neutron balance equation, which moves neutrons produced by delayed fission from their birth site to the precursors decay site. By incorporating these transport operators, the DNPs concentration is effectively removed from the balance equation, allowing the integral neutron balance equation to remain solely on the neutron flux. Precursors diffusion can also be accounted for by modeling the position increment of the delayed neutron as a Brownian motion. The position increment is sampled from a normal distribution with a variance proportional to the diffusion coefficient and the time-of-flight of the delayed neutron. The PDF of the position increment is derived and shown to be a solution of the 1D DNPs balance equation with constant coefficients.

The method was implemented in a one-velocity Monte-Carlo transport code and tested on a simplified model known as the rod problem. The results from the Monte-Carlo simulation were compared to those obtained from a deterministic calculation using identical parameters derived from the diffusion equation. The Monte-Carlo results were in agreement with the deterministic approach. Both implementations calculated neutron flux and reactivity as functions of the advection-reaction number \(\mathcal{B}\), with results aligning well with previous studies. As fuel velocity increased, the neutron flux was observed to shift, with the peak flux moving toward the core outlet. This shift is attributed to the drift of DNPs by the velocity field, causing delayed neutrons to be emitted away from the original fission site. The observed neutron flux shift was consistent with the deterministic calculation, confirming that the Monte-Carlo method accurately captures the impact of DNPs drift on neutron flux.

The reactivity difference as a function of fuel velocity was also calculated, showing consistency with the deterministic calculation. As observed in previous studies, the reactivity difference for large advection-reaction numbers was dependent on the size of the recirculation loop. Longer recirculation loops led to a greater reactivity difference, as more DNPs decayed in this section of the core. The same study was conducted with the addition of a diffusion term to the DNPs balance equation. Reactivity increased and then decreased as the diffusion-reaction number \(\mathcal{C}\) increased. This effect was linked to diffusion counteracting the drift of DNPs, increasing their number in the core and thus increasing reactivity.

This study demonstrated that DNPs drift by advection-diffusion can be effectively incorporated into Monte-Carlo simulations by simulating the motion of DNPs as a Brownian motion with a drift. It was shown to be equivalent in 1D to sampling the position of the DNPs from the Green's function of the DNPs balance equation. Future work will focus on extending this method to 2D systems and proving the equivalence between the Brownian motion with a drift and the Green's function of the DNPs balance equation in higher dimensions.
\section*{Additional Material}
The code used for the Monte-Carlo calculation is available on the author's GitHub page at \url{https://github.com/milliCoulomb/monte_carlo_rod}.
\section*{Acknowledgments}
Mathis Caprais would like to thank Cheikh Diop and Axel Fauvel for the fruitful discussions on Monte-Carlo methods. Special thanks to Aldo Dall'Osso for his careful review of the manuscript.
\bibliographystyle{unsrt}
\bibliography{bib.bib}
\end{document}